\documentclass{webofc}
\usepackage[varg]{txfonts}   % Web of Conferences font
\begin{document}
\title{Hadronic resonance production with ALICE at the LHC}
%
% subtitle is optional
%
%%%\subtitle{Do you have a subtitle?\\ If so, write it here}

\author{\firstname{Sergey} \lastname{Kiselev} for the ALICE collaboration\inst{1}\fnsep\thanks{\email{Sergey.Kiselev@cern.ch}} 
}

\institute{Institute for Theoretical and Experimental Physics, 117218 Moscow, Russia
          }

\abstract{%

We present recent results on short-lived hadronic resonances obtained 
by the ALICE experiment at LHC energies, including results from the Xe--Xe run.
The ALICE results on transverse momentum spectra, yields, their ratio to 
long-lived particles, and nuclear modification factors will be discussed. 
The results will be compared with model predictions and measurements at lower energies.
}
\maketitle
%

%Hadronic resonance production plays an important role both in elementary proton-proton and in heavy-ion collisions.
%The study of hadronic resonance production plays an important role in understanding the properties of the medium formed in heavy-ion collisions
Hadronic resonances are an important probe for the properties of the medium formed in heavy-ion collisions.
In heavy-ion collisions, since the lifetimes of short-lived resonances are comparable with
the lifetime of the late hadronic phase, regeneration and rescattering effects become
important and resonance ratios to longer lived particles can be used to estimate
the time interval between the chemical and kinetic freeze-out~\cite{Timescale}.
The measurements in pp and p-Pb collisions constitute a reference for nuclear collisions
and provide information for tuning event generators inspired by Quantum Chromodynamics.
% study of collectivity in small systems.

Recent results on short-lived mesonic $\rho(770)^{0}$, $\mathrm{K}^{*}(892)^{0}$, $\phi(1020)$ as well as baryonic $\Lambda(1520)$ and 
$\Xi(1530)^{0}$ resonances (hereafter $\rho^{0}$, $\mathrm{K}^{*0}$, $\phi$, $\Lambda^{*}$, $\Xi^{*0}$) 
obtained by the ALICE experiment are presented.
The $\rho^{0}$ has been measured in pp and Pb--Pb collisions at $\sqrt{s_{\rm NN}}$ = 2.76 TeV and published recently in~\cite{ALICErho0}.
The $\mathrm{K}^{*0}$ and $\phi$ have been measured in pp collisions at $\sqrt{s}$ = 13 TeV, in Xe--Xe collisions at $\sqrt{s_{\rm NN}}$ = 5.44 TeV and in Pb--Pb collisions at $\sqrt{s_{\rm NN}}$ = 5.02 TeV
(results for the $\mathrm{K}^{*0}$ and $\phi$ in pp at $\sqrt{s}$ = 7 TeV, p-Pb at $\sqrt{s_{\rm NN}}$~=~5.02~TeV
and Pb--Pb at $\sqrt{s_{\rm NN}}$ = 2.76 TeV published in~\cite{ALICEpp7}, ~\cite{ALICEpPb} and  ~\cite{{ALICEPbPb}, {ALICEPbPb-highPT}}, respectively).
The $\Lambda^{*}$ has been measured in Pb--Pb collisions at $\sqrt{s_{\rm NN}}$ = 2.76 TeV and published recently in~\cite{ALICELambdaStar}. 
The $\Xi^{*0}$ has been measured in Pb--Pb collisions at $\sqrt{s_{\rm NN}}$~=~2.76~TeV (results for the $\Sigma^{*\pm}$ and $\Xi^{*0}$ in pp at $\sqrt{s}$ = 7 TeV and p-Pb at $\sqrt{s_{\rm NN}}$~=~5.02~TeV published in~\cite{ALICEppSigmaStar} and ~\cite{ALICEpPbSigmaStar}, respectively).
%In different multiplicity or centrality intervals.

The  resonances  are  reconstructed  in  their  hadronic  decay  channels  and have very different lifetimes
as shown in Table~\ref{tab:Res}. 
\begin{table}[ht]
\begin{center}
%\begin{tabular}{|c|c|c|c|c|c|}
\begin{tabular}{ c c c c c c }
\hline
                    &$\rho^{0}$&$\mathrm{K}^{*0}$  &  $\phi$ &    $\Lambda^{*}$    &     $\Xi^{*0}$      \\
\hline
%decay channel      &$\pi^{\pm} + \mathrm{K}^{\mp}$&$\mathrm{K}^{+} + \mathrm{K}^{-}$&$p(\bar{p}) + {K}^{\mp}$&$\Xi^{-} + \pi^{+}
decay channel (B.R.)&$\pi\pi (1.00) $&$\mathrm{K}\pi (0.67) $&$\mathrm{K}\mathrm{K} (0.49) $&$p\mathrm{K} (0.22)$&$\Xi\pi (0.67)$\\
%\hline
lifetime (fm/\it{c})& 1.3 & 4.2 & 46.2 & 12.6 & 21.7 \\ 
\hline
\end{tabular}
\end{center}
\caption{Reconstructed decay mode, branching ratio and lifetime values for hadronic resonances}
 \label{tab:Res}
\end{table}

\begin{figure}[hbtp]
\begin{center}
\includegraphics[scale=0.32]{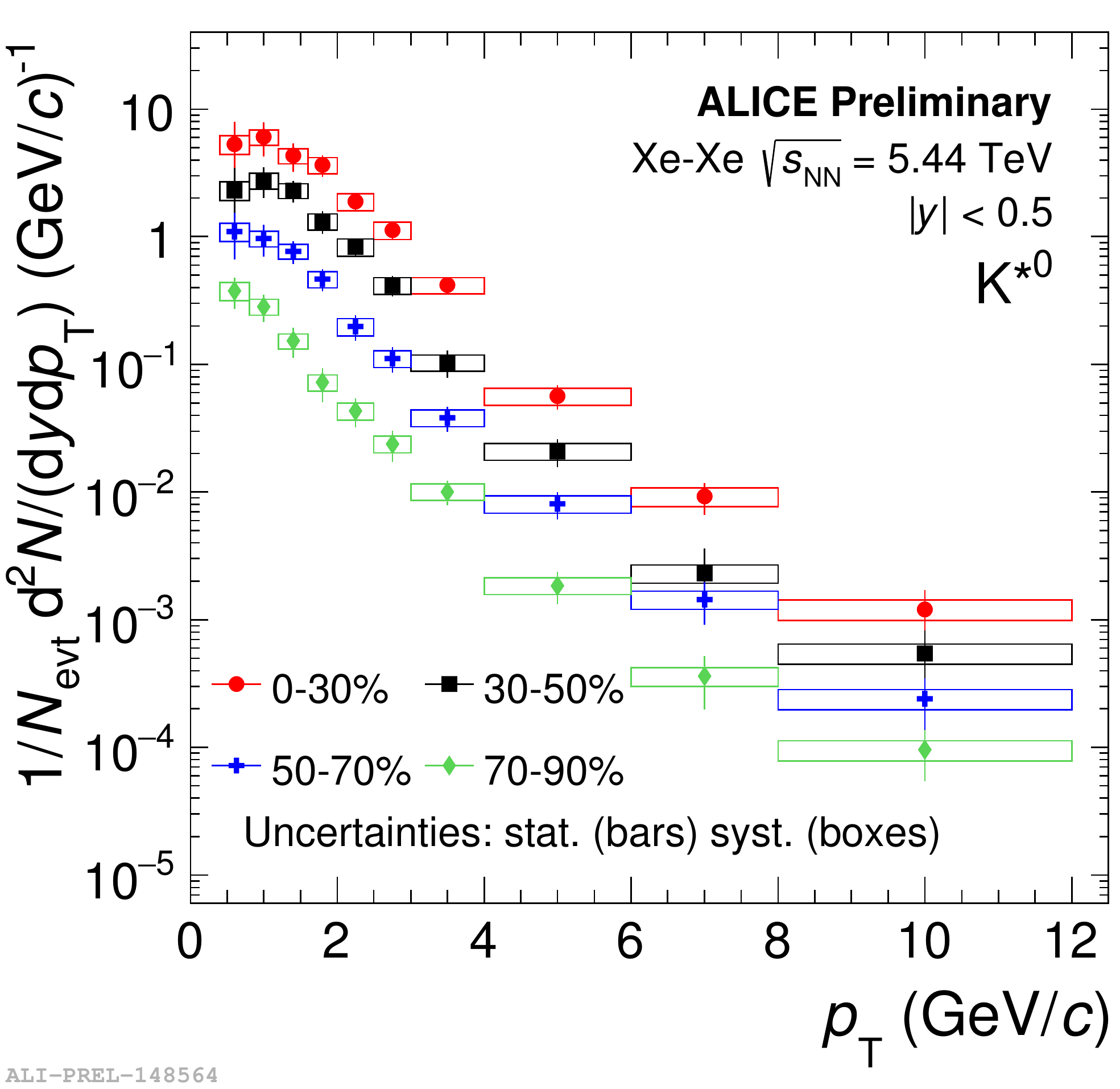}
\includegraphics[scale=0.32]{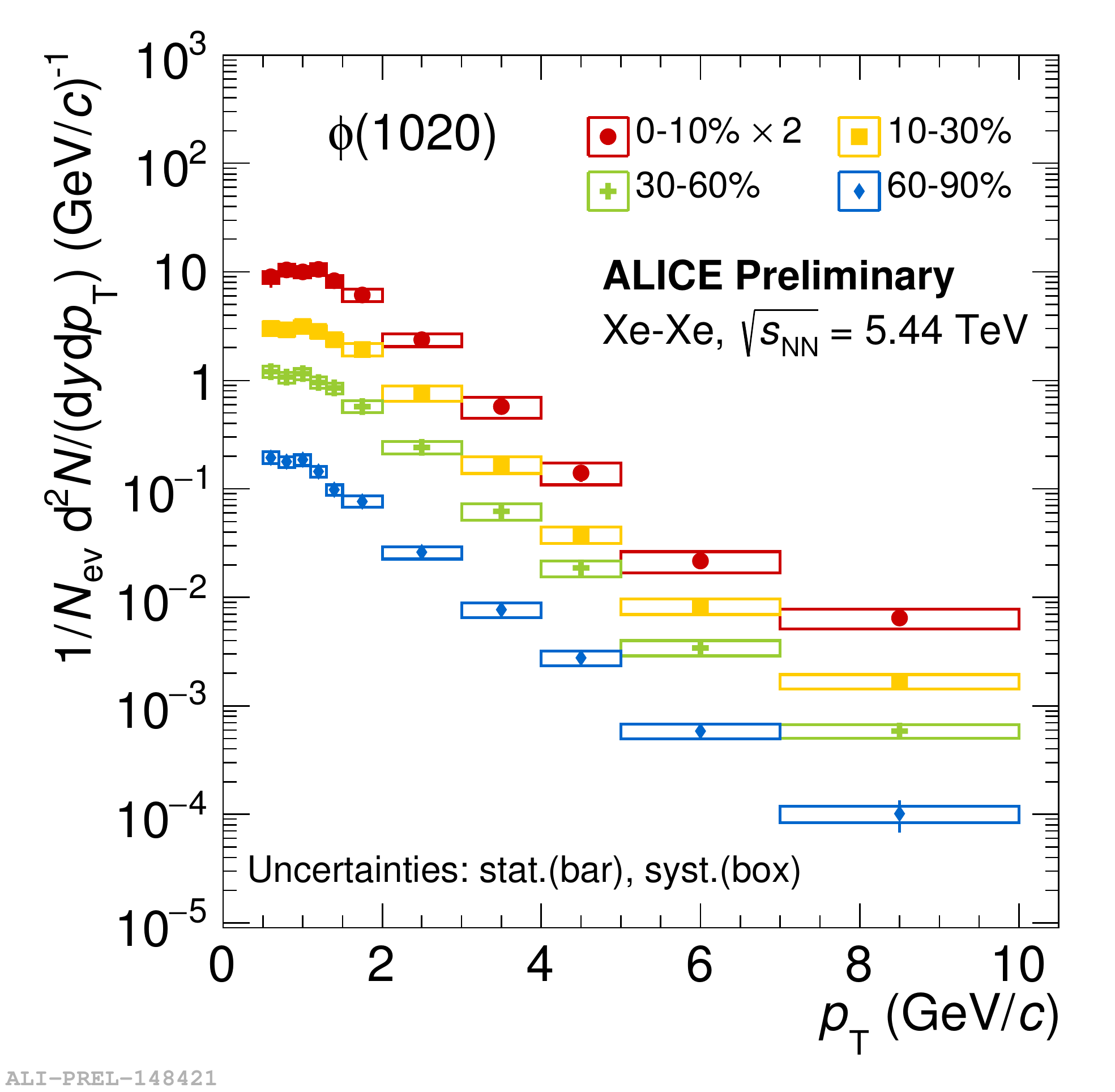}
\end{center}
\caption{(color online) Transverse momentum spectra for $\mathrm{K}^{*0}$ (left) and $\phi$ (right) in different centrality classes of Xe--Xe collisions at $\sqrt{s_{NN}}$ = 5.44 TeV.
}
  \label{fig:spectra}
\end{figure}

Figure~\ref{fig:spectra} presents the transverse momentum spectra for $\mathrm{K}^{*0}$ and $\phi$ in Xe--Xe collisions at $\sqrt{s_{NN}}$ = 5.44 TeV.
The spectra have been measured for different centrality up to $p_\mathrm{T} = 10$ GeV/$c$.

\begin{figure}[hbtp]
\begin{center}
\includegraphics[scale=0.20]{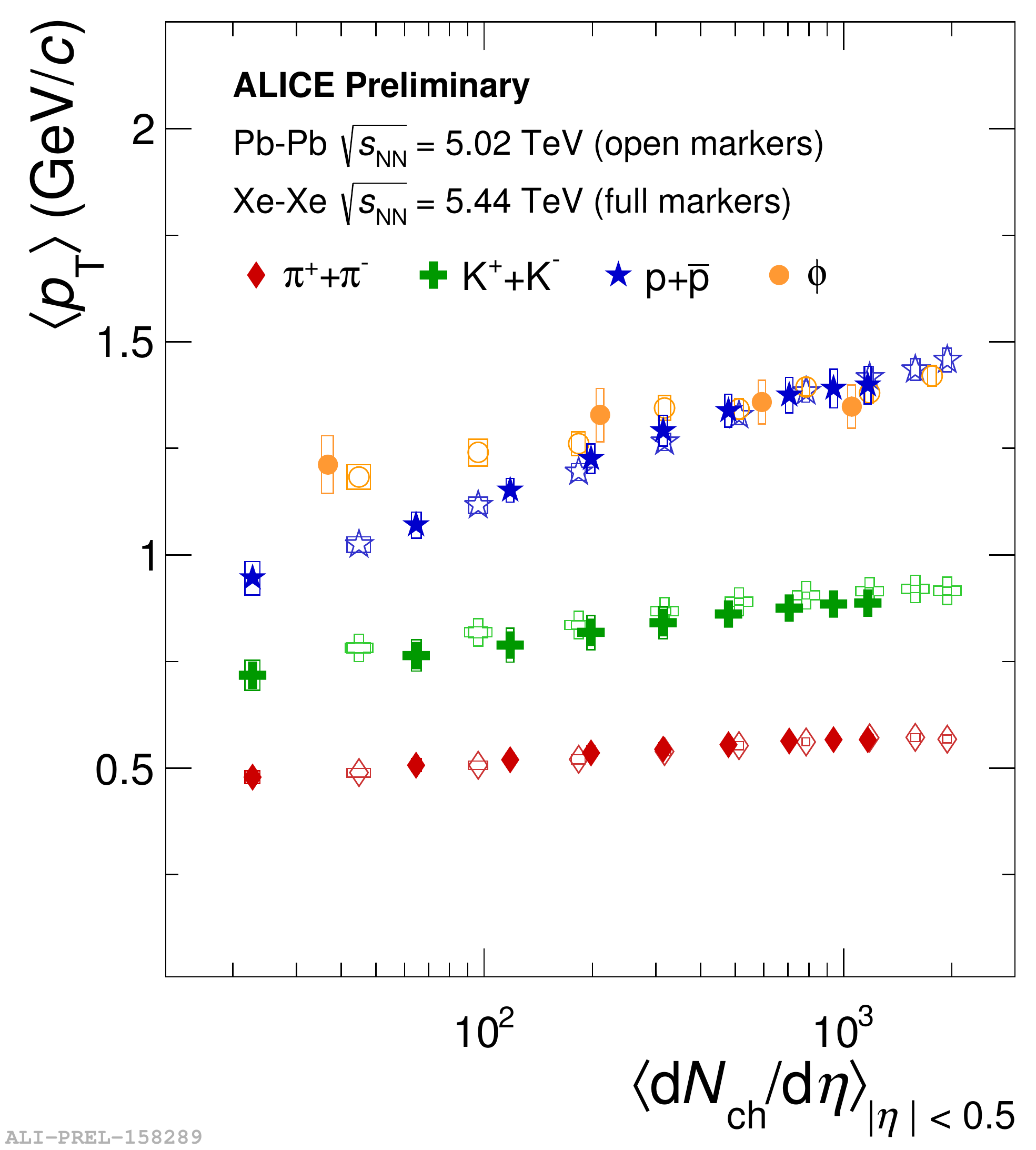}
\includegraphics[scale=0.44]{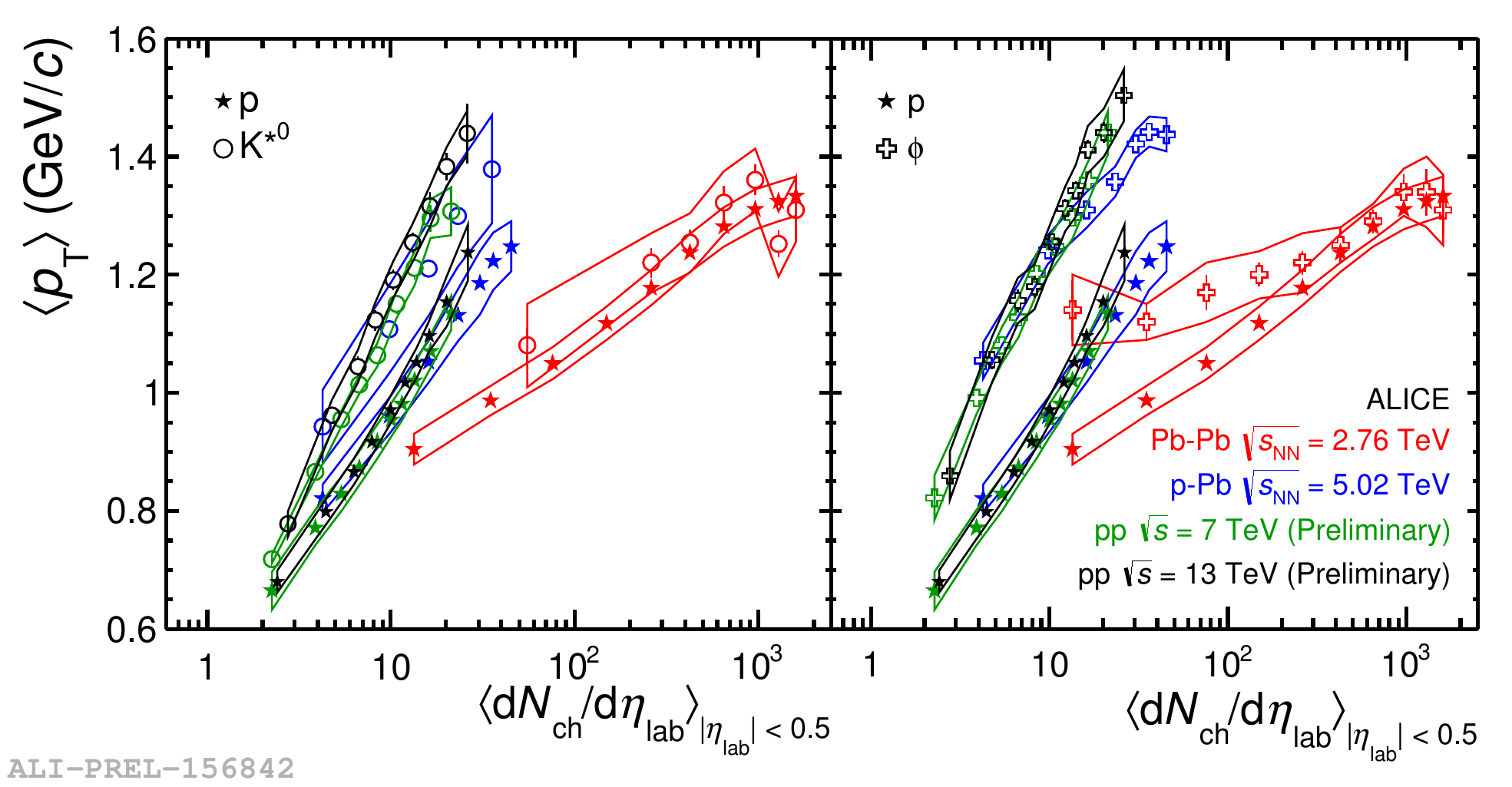}
\end{center}
\caption{(color online) 
The mean transverse momentum as a function of the charged-particle multiplicity density. 
(left) pion, kaon, proton and $\phi$.
(right) $\mathrm{K}^{*0}$ (left panel) and $\phi$ (right panel) compared to p (same data in both panels).
%Results for p-Pb and Pb--Pb systems are from ~\cite{ALICEpPb} and ~\cite{{ALICEPbPb}, respectively.
}
  \label{fig:mpt}
\end{figure}

The mean transverse momenta of $\phi$ and stable hadrons in Xe--Xe collisions as a function of the charged-particle multiplicity density are shown in Fig.~\ref{fig:mpt} (left).
Results for Xe--Xe collisions confirm the trends observed in Pb--Pb collisions.
In central collisions we observe mass ordering.
The $\phi$ and p, which have similar masses, are observed to have
similar $\langle p_\mathrm{T}\rangle$ values, as expected if their spectral shape is dominated by radial flow.

Figure~\ref{fig:mpt} (right) presents the $\langle p_\mathrm{T}\rangle$ of $\mathrm{K}^{*0}$, $\phi$ and p 
as a function of the charged-particle multiplicity density measured for pp collisions at $\sqrt{s}$ = 13 TeV
and compared with the results obtained in pp, p–Pb~\cite{ALICEpPb} and Pb–Pb~\cite{ALICEPbPb} collisions at $\sqrt{s_{NN}}$ = 7, 5.02 and 2.76 TeV, respectively. In pp collisions the $\langle p_\mathrm{T}\rangle$ increase with multiplicity at $\sqrt{s}$ = 13 TeV is similar to the one at $\sqrt{s}$ = 7 TeV.
The $\langle p_\mathrm{T}\rangle$  values follow a similar trend 
with multiplicity
%the system size 
in pp and p--Pb collisions, where they rise faster with multiplicity than in Pb--Pb collisions.
An analogous behavior has been observed in~\cite{ALICE_mpt} for charged particles and can be understood as the effect of color reconnection between strings produced in multi-parton interactions. 
In central Pb–Pb collisions, the $\langle p_\mathrm{T}\rangle$ values for these three particles are consistent within uncertainties, as it is expected in presence of a common radial flow.
However in pp and p--Pb collisions the $\langle p_\mathrm{T}\rangle$ values for the $\mathrm{K}^{*0}$ and $\phi$ resonances are higher than for p.
The mass ordering observed in central Pb–Pb collisions, where particles with similar mass have
similar $\langle p_\mathrm{T}\rangle$, is not observed in pp and p--Pb collisions~\cite{ALICEpPb}.

\begin{figure}[hbtp]
\begin{center}
\includegraphics[scale=0.48]{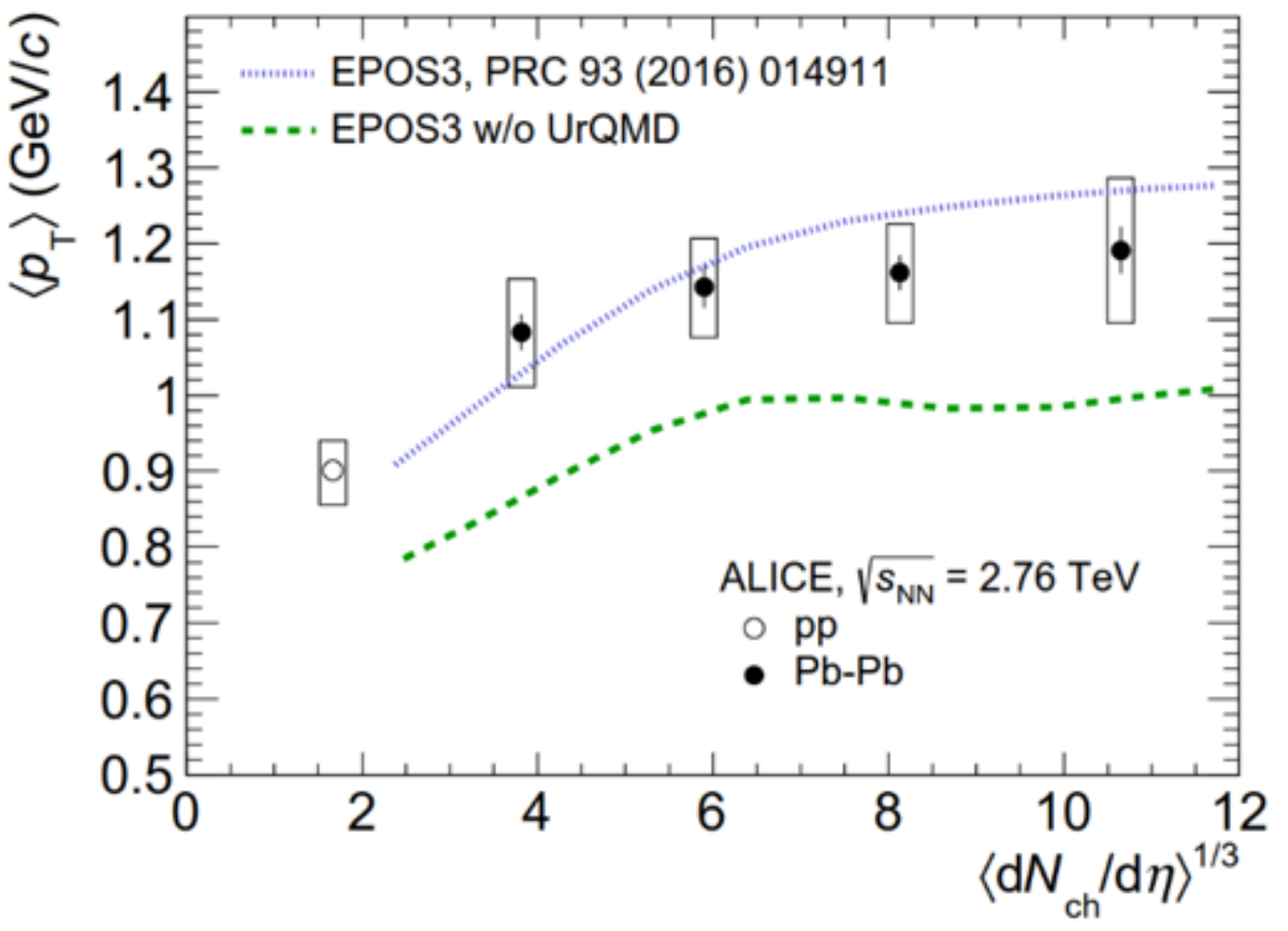}
\includegraphics[scale=0.26]{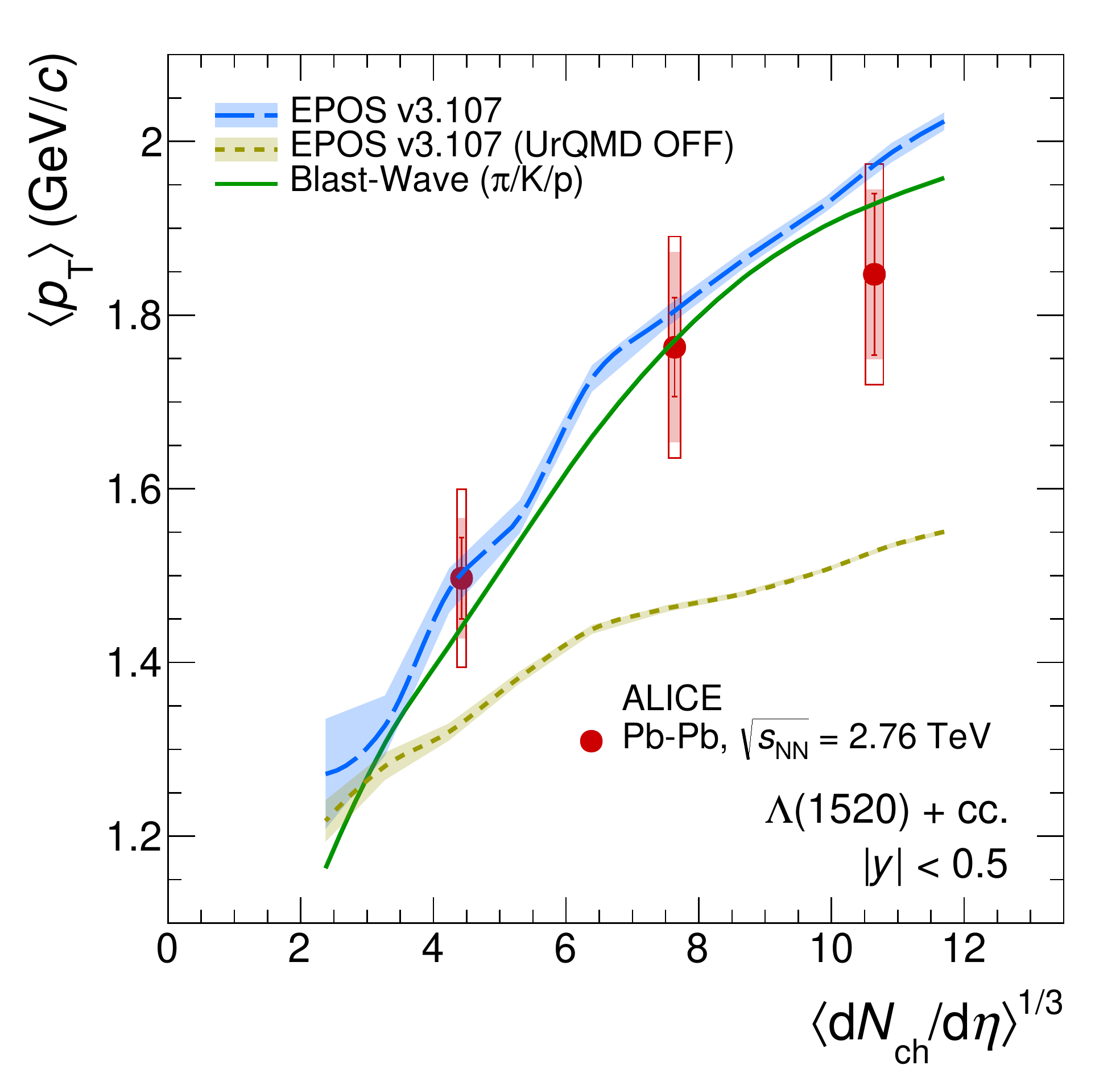}
\end{center}
\caption{(color online) 
The mean transverse momentum of $\rho^{0}$ (left)~\cite{ALICErho0} and $\Lambda^{*}$ (right)~\cite{ALICELambdaStar} as a function of the charged-particle multiplicity density. The measurements are also compared to model predictions: 
EPOS3~\cite{KnospeEPOS}, Blast-Wave~\cite{BlastWave}.
}
  \label{fig:mptPbPb}
\end{figure}

Figure~\ref{fig:mptPbPb} shows the $\langle p_\mathrm{T}\rangle$ as a function of the charged-particle multiplicity density
for $\rho^{0}$~\cite{ALICErho0} and $\Lambda^{*}$~\cite{ALICELambdaStar} in Pb--Pb collisions at $\sqrt{s_{NN}}$~=~2.76~TeV. 
The results are in agreement with the prediction from the EPOS3 generator with 
UrQMD~\cite{KnospeEPOS}, which includes a modeling of re-scattering and regeneration in the hadronic phase.
The results for $\Lambda^{*}$ are also in agreement with the average momentum extracted from the Blast-Wave model~\cite{BlastWave} 
with parameters obtained from the simultaneous fit to pion, kaon, and (anti)proton $p_{T}$ distributions~\cite{BlastWavePbPb}.

\begin{figure}[hbtp]
\begin{center}
\includegraphics[scale=0.3]{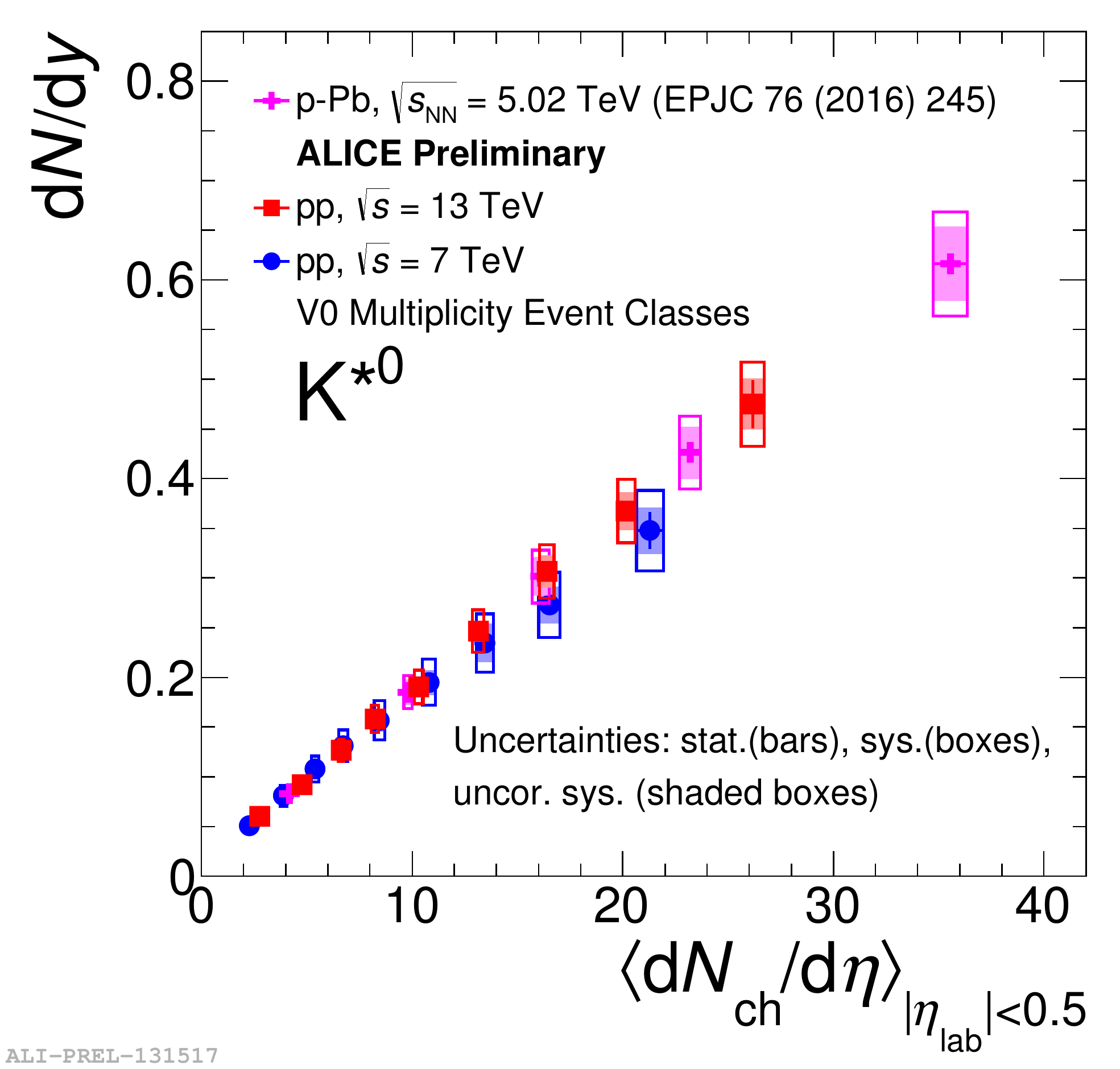}
\includegraphics[scale=0.3]{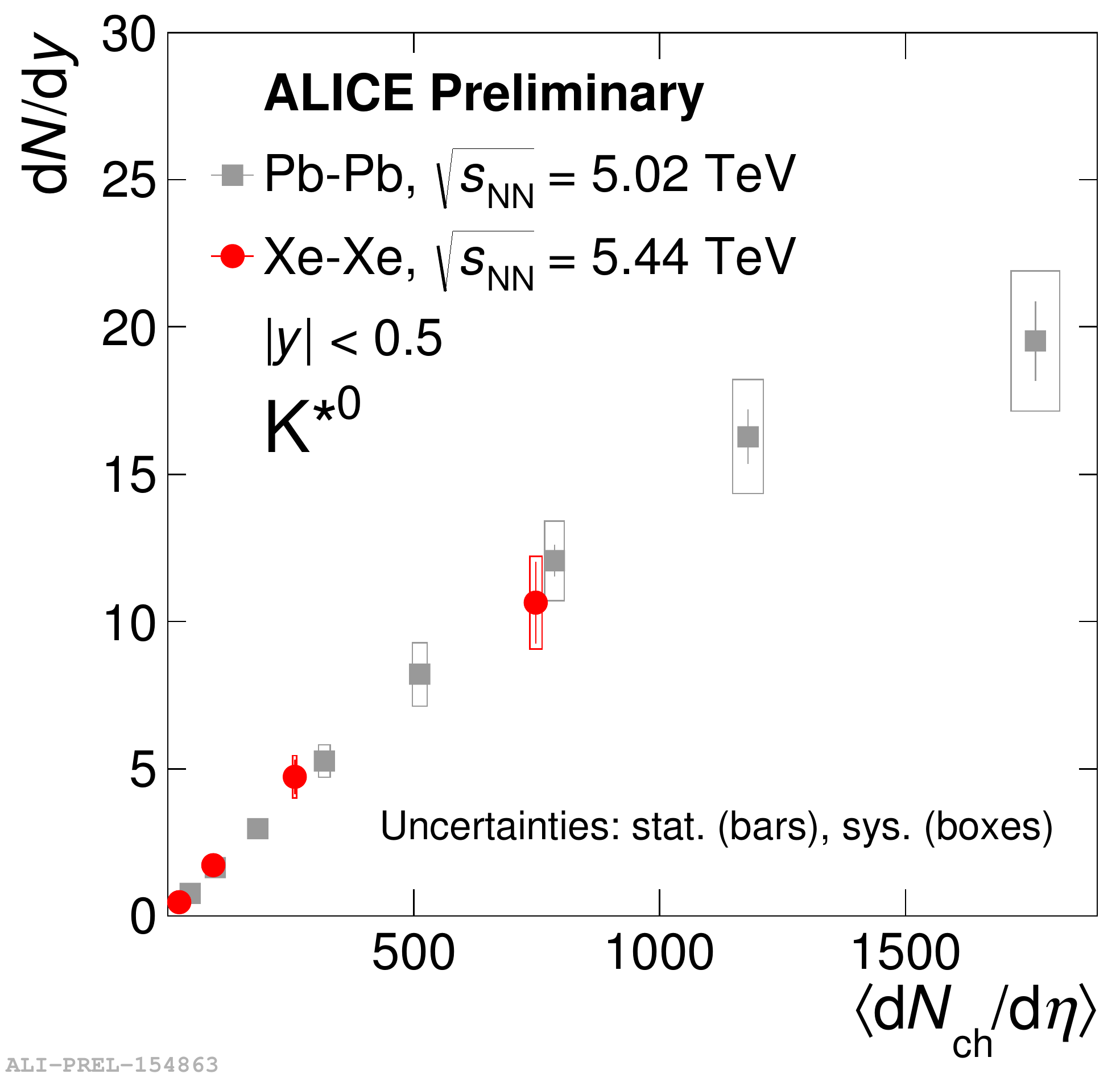}
\includegraphics[scale=0.3]{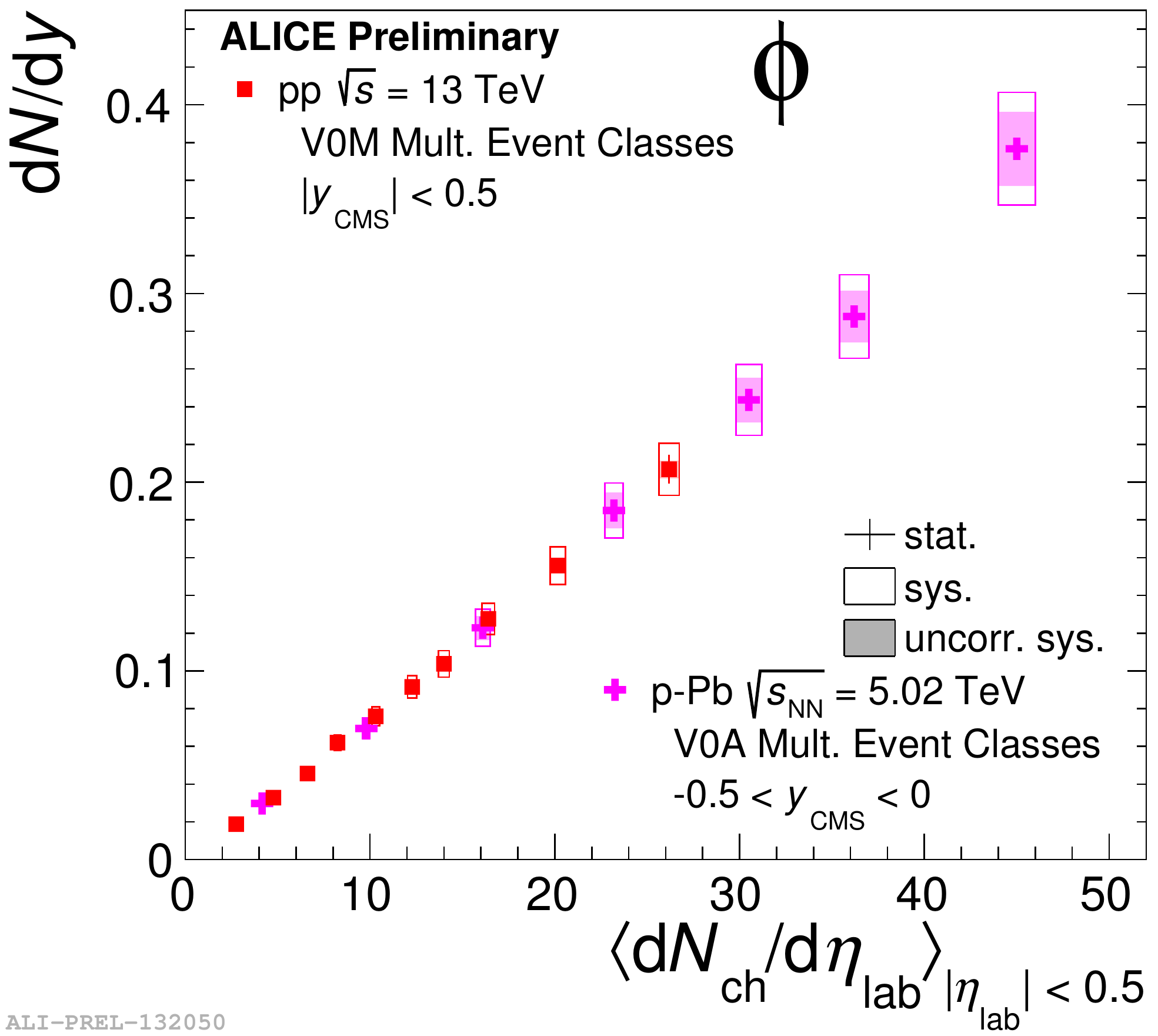}
\includegraphics[scale=0.3]{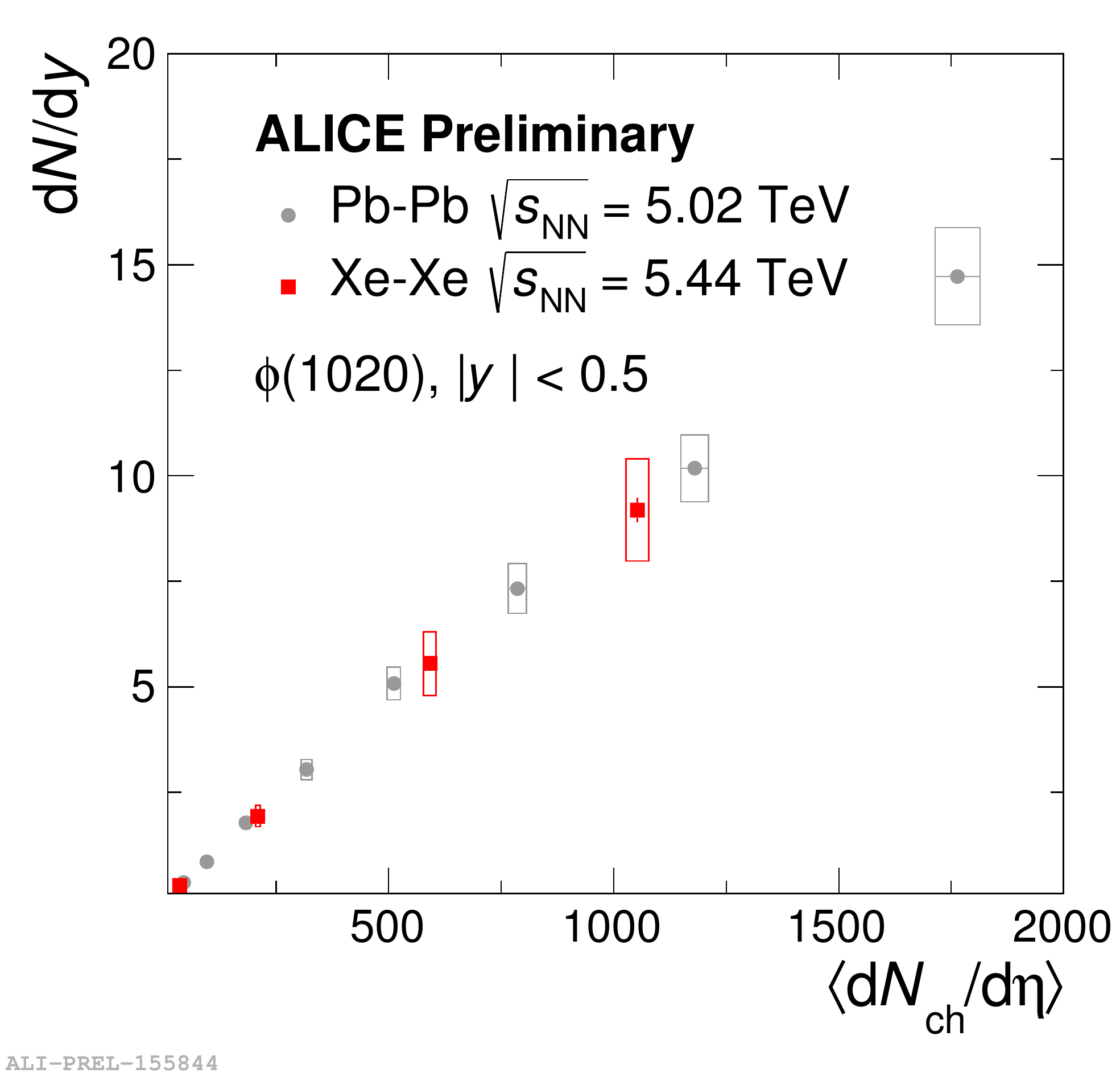}
\end{center}
\caption{(color online) $p_{T}$-integrated yields (dN/dy) of $\mathrm{K}^{*0}$ (top) and $\phi$ (bottom)
 as a function of the charged-particle multiplicity density in pp and p-Pb~\cite{ALICEpPb} (left) and Xe--Xe and Pb--Pb (right) collisions. 
}
  \label{fig:yields}
\end{figure}

Figure~\ref{fig:yields} presents $p_{T}$-integrated yields of $\mathrm{K}^{*0}$ and $\phi$ 
in pp, p-Pb~\cite{ALICEpPb}, Xe--Xe and Pb--Pb collisions as a function of the charged-particle multiplicity density.
Yields are independent of collision system and appear to be driven by event multiplicity.

\begin{figure}[hbtp]
\begin{center}
\includegraphics[scale=0.25]{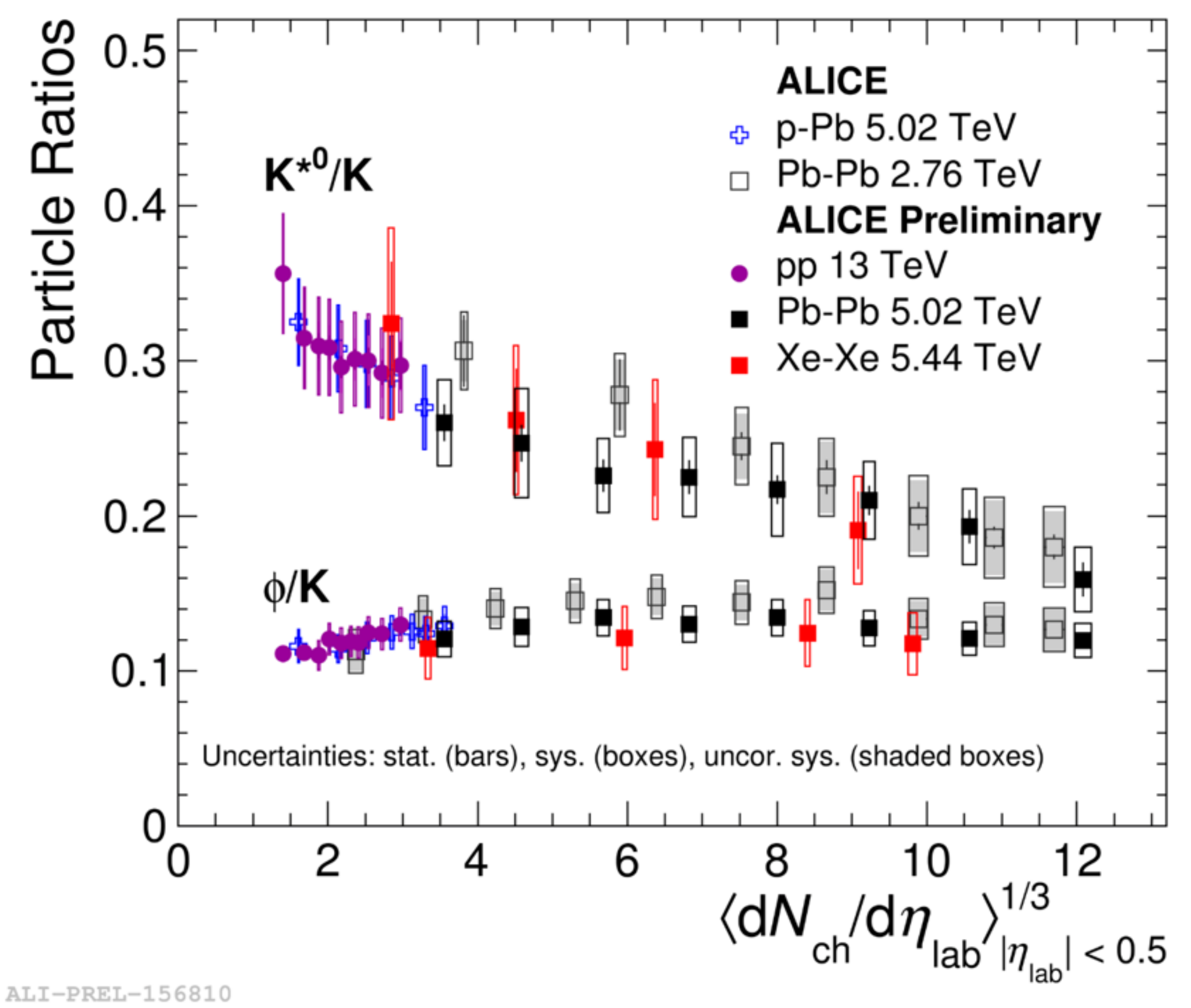}
\includegraphics[scale=0.29]{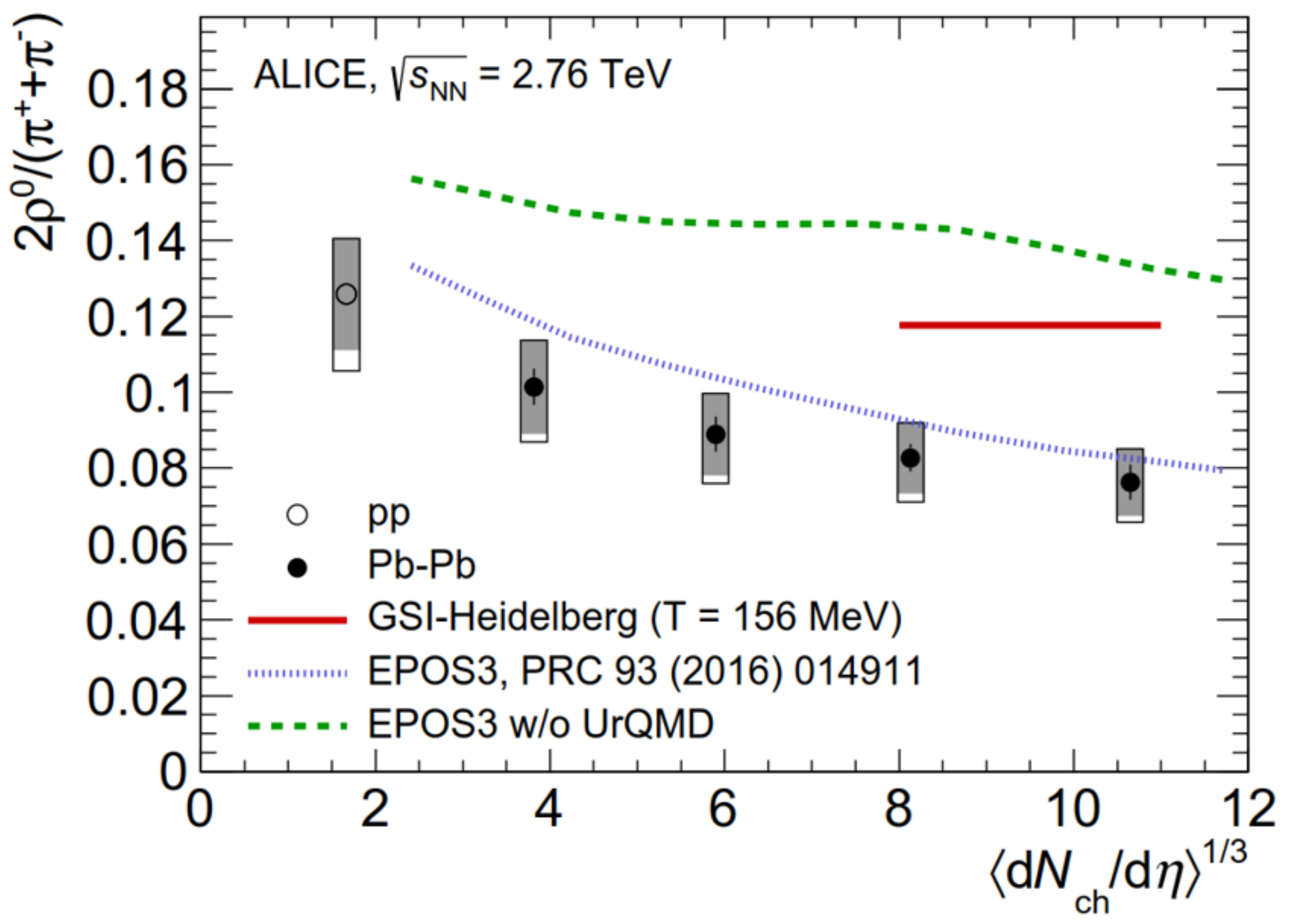}
\includegraphics[scale=0.29]{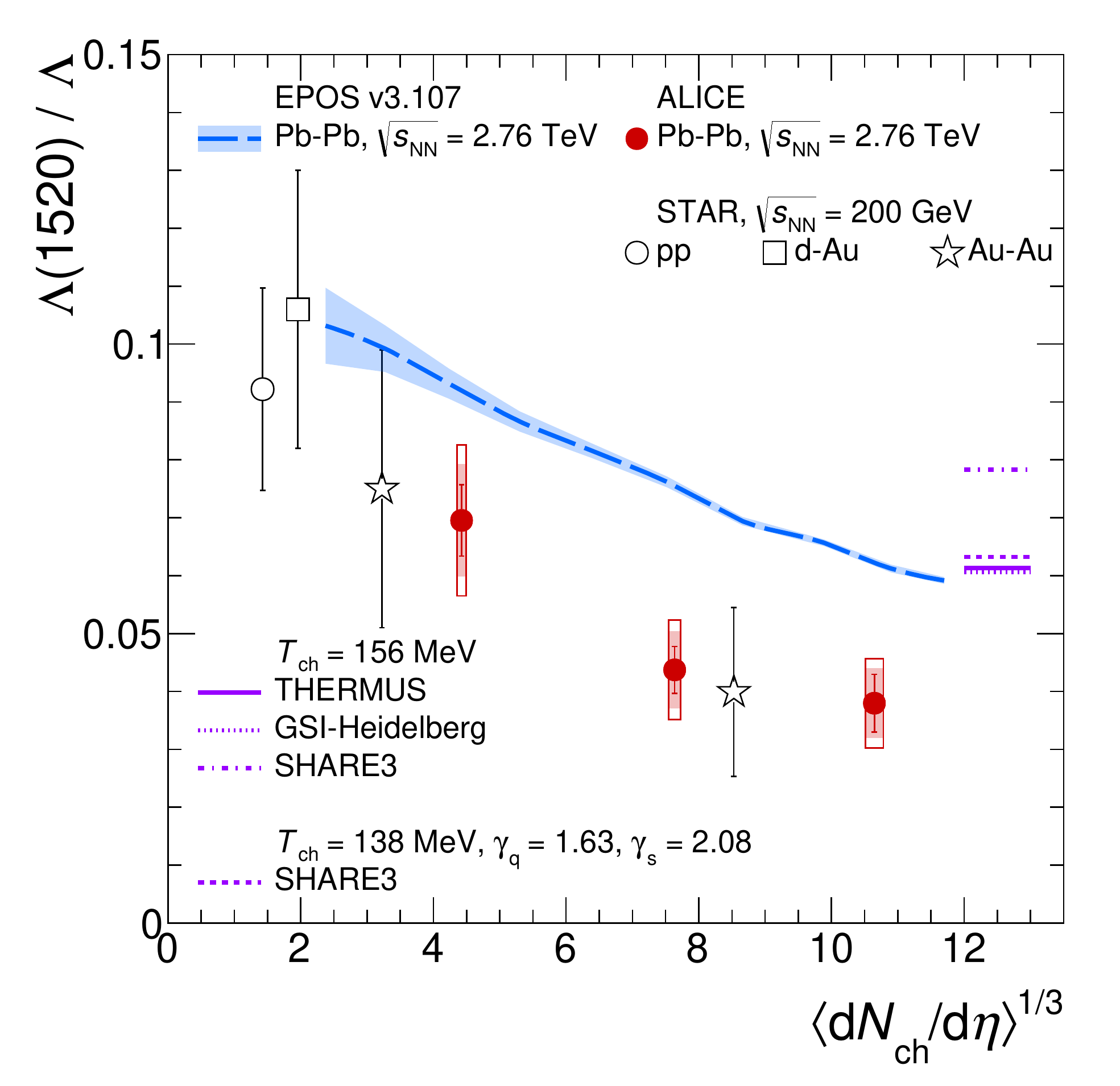}
\includegraphics[scale=0.29]{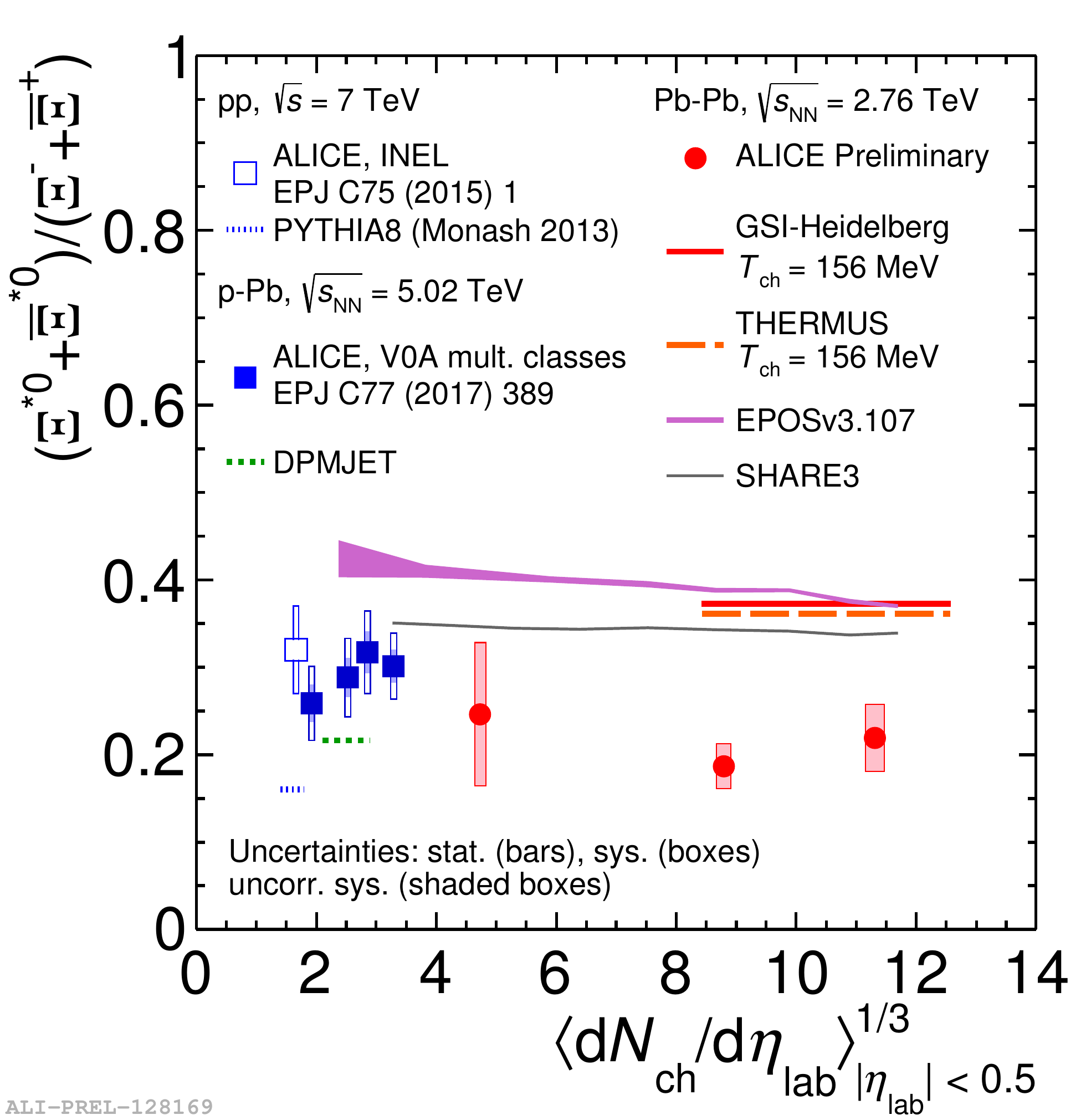}
\end{center}
\caption{(color online) Particle yield ratios $\mathrm{K}^{*0}/\mathrm{K}$ 
and $\phi/\mathrm{K}$ (top left), $\rho^{0}/\pi$ (top right)~\cite{ALICErho0}, $\Lambda^{*}/\Lambda$ (bottom left)~\cite{ALICELambdaStar} and $\Xi^{*0}/\Xi$ (bottom right) as a function of the charged-particle multiplicity density for various collision systems. STAR data from~\cite{STAR}. The measurements are also compared to model predictions: 
EPOS3~\cite{KnospeEPOS}, THERMUS~\cite{THERMUS}, GSI-Helderberg ~\cite{GSIHelderberg}, SHARE~\cite{SHARE}.  
}
  \label{fig:Mratios}
\end{figure}

Figure~\ref{fig:Mratios} (top left) shows the particle yield ratios $\mathrm{K}^{*0}/\mathrm{K}$ and $\phi/\mathrm{K}$ 
in Xe--Xe collisions at $\sqrt{s_{\rm NN}}$~=~5.44~TeV. 
Results for p-Pb  collisions at $\sqrt{s_{\rm NN}}$~=~5.02~TeV~\cite{ALICEpPb}, Pb--Pb collisions at $\sqrt{s_{\rm NN}}$~=~2.76~TeV~\cite{ALICEPbPb-highPT}, pp collisions at $\sqrt{s}$ = 13 TeV and Pb--Pb collisions at $\sqrt{s_{\rm NN}}$ = 5.02 TeV are also shown. 
%Results for Xe--Xe collisions confirm the trends observed in Pb--Pb collisions.
The $\mathrm{K}^{*0}/\mathrm{K}$ ratio shows a significant suppression going from p-Pb and peripheral Xe--Xe/Pb--Pb collisions 
to most central Xe--Xe/Pb--Pb collisions. This suppression is consistent with rescattering of $\mathrm{K}^{*0}$ daughters in the hadronic phase of central collisions as the dominant effect.
There is a hint of decrease of the ratio with increasing multiplicity in pp and p-Pb collisions.
The values of the ratio are consistent for similar multiplicities across collision systems (pp, p-Pb).
The decrease of the ratio might be an indication of a hadron-gas phase with non-zero lifetime in high-multiplicity pp and p-Pb collisions.
The $\phi/\mathrm{K}$ ratio is nearly flat.
% and agrees with the prediction of the thermal model. 
This suggests that rescattering effects are not important for $\phi$, which has 10 times longer lifetime 
than $\mathrm{K}^{*0}$ and decays mainly after the kinetic freeze-out.
In Pb--Pb collisions at $\sqrt{s_{\rm NN}}$~=~2.76~TeV we observe $\rho^{0}/\pi$~\cite{ALICErho0} and $\Lambda^{*}/\Lambda$~\cite{ALICELambdaStar}
ratio suppression similar to the $\mathrm{K}^{*0}/\mathrm{K}$ ratio, 
Fig.~\ref{fig:Mratios} (top right) and Fig.~\ref{fig:Mratios} (bottom left), respectively.
The $\Lambda^{*}/\Lambda$ suppression confirms the trend seen by STAR at $\sqrt{s_{\rm NN}}$~=~200~GeV~\cite{STAR}.
For the $\Xi^{*0}/\Xi$ ratio, Fig.~\ref{fig:Mratios} (bottom right), there is a hint of suppression, 
%in central Pb--Pb collisions with respect to pp and p-Pb collisions, 
but systematics are to be reduced in peripheral Pb--Pb collisions before making any conclusive statement.
Although predictions of the EPOS3 model with UrQMD~\cite{KnospeEPOS} overestimate the data, 
the trend of the suppression is qualitatively reproduced for $\mathrm{K}^{*0}$, $\rho^{0}$, and $\Lambda^{*}$.
Thermal model predictions ~\cite{{THERMUS}, {GSIHelderberg}, {SHARE}} overestimate all particle ratios under study in central Pb--Pb collisions, 
except the $\phi/\mathrm{K}$ ratio.

\begin{figure}[hbtp]
\begin{center}
\includegraphics[scale=0.39]{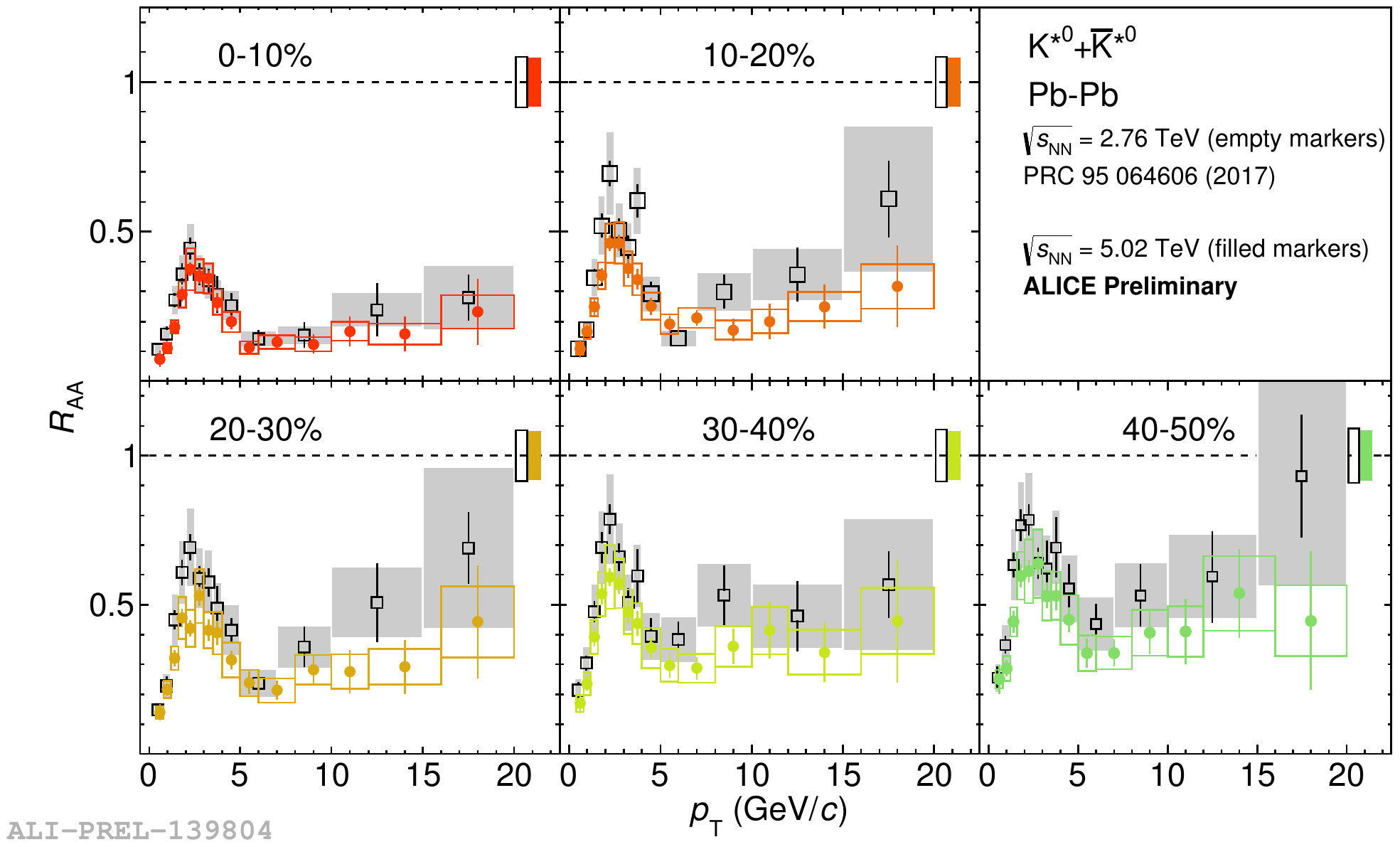}
\includegraphics[scale=0.24]{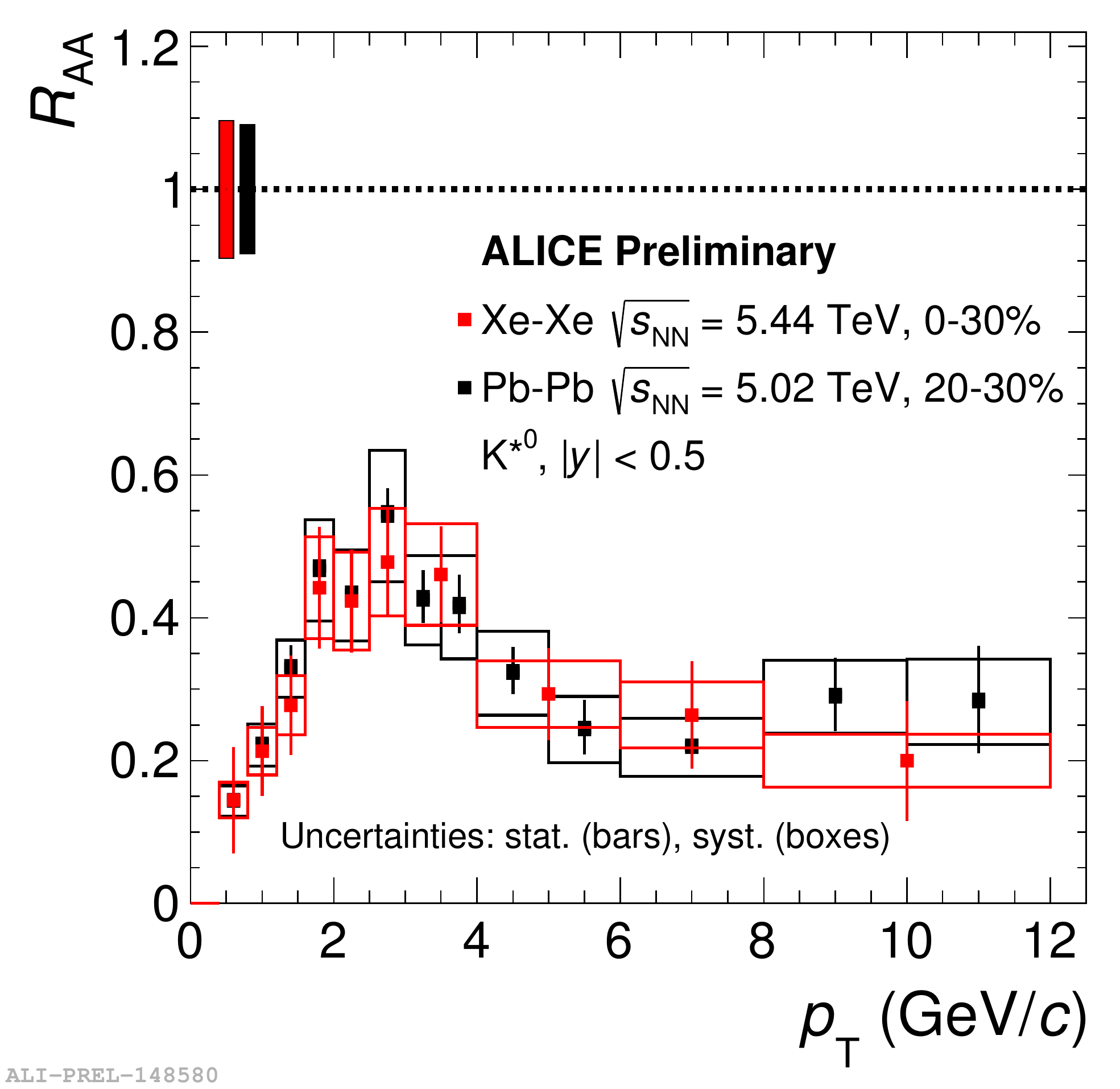}
\end{center}
\caption{(color online) The nuclear modification factor $R_{AA}$ for $\mathrm{K}^{*0}$ as a function of transverse momentum.
(left)  Pb--Pb collisions at $\sqrt{s_\mathrm{NN}}$ = 2.76 TeV~\cite{ALICEPbPb-highPT} and $\sqrt{s_\mathrm{NN}}$ = 5.02 TeV in different centrality. 
(right)  Xe--Xe and Pb--Pb collisions at the same multiplicity (and not just centrality percentile). 
%(right) $\mathrm{K}^{*0}$, $\phi$ and stable hadrons in different centrality of Pb--Pb collisions 
%at $\sqrt{s_\mathrm{NN}}$ = 5.02 TeV. 
}
  \label{fig:RAA}
\end{figure}

In central Pb--Pb collisions  the nuclear modification factor $R_{AA}$ for 
$\mathrm{K}^{*0}$~\cite{ALICEPbPb-highPT}, $\phi$~\cite{ALICEPbPb-highPT} and $\rho^{0}$~\cite{ALICErho0} 
is consistent with light-flavored hadrons at $p_\mathrm{T} > 8$ GeV/$c$, demonstrating strong suppression. 
%$R_{AA} \approx$ 0.15 - 0.2.
At low $p_\mathrm{T}$, $p_\mathrm{T} < 2$ GeV/$c$, the $\mathrm{K}^{*0}$ and $\rho^{0}$ are more suppressed than light-flavored hadrons,
which is consistent with the hypothesis that flow and rescattering effects are important.
Figure~\ref{fig:RAA} (left) presents $R_{AA}$ of $\mathrm{K}^{*0}$ for Pb--Pb collisions 
at $\sqrt{s_\mathrm{NN}}$ = 2.76 TeV~\cite{ALICEPbPb-highPT} and $\sqrt{s_\mathrm{NN}}$ = 5.02 TeV in different centrality.
We do not observe significant energy dependence of $R_{AA}$.
$R_{AA}$ of $\mathrm{K}^{*0}$ in Xe--Xe and Pb--Pb collisions are consistent within uncertainties once compared
at the same multiplicity (and not just centrality percentile) as shown in Fig.~\ref{fig:RAA} (right).

In summary, recent results on short-lived hadronic resonances obtained by the ALICE experiment in pp, p-Pb, Xe--Xe and Pb--Pb collisions at the LHC energies have been presented.
In pp and p--Pb collisions the $\langle p_\mathrm{T}\rangle$ values for the $\mathrm{K}^{*0}$ and $\phi$ resonances 
rise faster with multiplicity than in Pb–Pb collisions.
One possible explanation could be the effect of color reconnection between strings produced in multi-parton interactions.
The mass ordering observed in central Xe--Xe and Pb–Pb collisions, where particles with similar masses ($\mathrm{K}^{*0}$, $\phi$ and p) have similar $\langle p_\mathrm{T}\rangle$, is not observed in pp and p--Pb collisions.
Yields of $\mathrm{K}^{*0}$ and $\phi$ in pp, p-Pb, Xe--Xe and Pb--Pb collisions are independent of collision system 
and appear to be driven by event multiplicity.
The  $\mathrm{K}^{*0}/\mathrm{K}$, $\rho^{0}/\pi$ and $\Lambda^{*}/\Lambda$ ratios exhibit a significant suppression 
going from peripheral to central Pb--Pb collisions, consistent with rescattering of the decay products of the short-lived resonances in the hadronic phase.
The $\phi/\mathrm{K}$ ratio is not suppressed due to the longer lifetime of the $\phi$.
Results for the $\mathrm{K}^{*0}/\mathrm{K}$ and $\phi/\mathrm{K}$ ratios in Xe--Xe collisions confirm the trend observed in Pb--Pb collisions.
There is a hint of suppression for the $\Xi^{*0}/\Xi$ ratio,  
but systematics are to be reduced in peripheral Pb--Pb collisions before making any conclusive statement.
Although predictions of the EPOS3 model with UrQMD overestimate the data, 
the trend of the suppression is qualitatively reproduced.
Thermal model predictions overestimate all particle ratios under study in central Pb--Pb collisions, 
except the $\phi/\mathrm{K}$ ratio.
In central Pb--Pb collisions  the nuclear modification factor $R_{AA}$ for 
$\mathrm{K}^{*0}$, $\phi$ and $\rho^{0}$ 
consistent with light-flavored hadrons at $p_\mathrm{T} > 8$ GeV/$c$.
At low $p_\mathrm{T}$, $p_\mathrm{T} < 2$ GeV/$c$, the $\mathrm{K}^{*0}$ and $\rho^{0}$ are more suppressed than light-flavored hadrons,
which would be consistent with the hypothesis that flow and rescattering effects are important.
We do not observe significant energy dependence of $R_{AA}$ in Pb--Pb collisions.
$R_{AA}$ of $\mathrm{K}^{*0}$ in Xe--Xe and Pb--Pb collisions are consistent within uncertainties once compared
at the same multiplicity.

\end{document}